\documentclass[runningheads]{llncs}
\usepackage{graphicx}
\usepackage{hyperref}
\usepackage{xcolor}
\usepackage[ruled,vlined]{algorithm2e}
\usepackage{multirow}
\usepackage{subcaption}
\usepackage{booktabs}
\usepackage{paralist}
\usepackage{enumerate}
\usepackage{enumitem}

\SetCommentSty{mycommfont}
\SetKwInput{KwInput}{Input}                
\SetKwInput{KwOutput}{Output}              

\newlength\mylen

\usepackage{amsmath}

\begin{document}
\title{Mutual Information Scoring: Increasing Interpretability in Categorical Clustering Tasks with Applications to Child Welfare Data}
\titlerunning{Mutual Information Scoring}
%
\author{Pranav Sankhe\inst{1} \and
Seventy F. Hall\inst{1} \and Melanie Sage\inst{1} \and Maria Y. Rodriguez\inst{1} \and Varun Chandola\inst{1} \and Kenneth Joseph\inst{1}}
\authorrunning{P. Sankhe et al.}

\institute{University at Buffalo, Buffalo, NY\\
\email{{pranavgi,sfhall,msage,myr2,chandola,kjoseph}@buffalo.edu}}
\maketitle              
\begin{abstract}
Youth in the American foster care system are significantly more likely than their peers to face a number of negative life outcomes, from homelessness to incarceration. Administrative data on these youth have the potential to provide insights that can help identify ways to improve their path towards a better life. However, such data also suffer from a variety of biases, from missing data to reflections of systemic inequality. The present work proposes a novel, prescriptive approach to using these data to provide insights about both data biases and the systems and youth they track. Specifically, we develop a novel categorical clustering and cluster summarization methodology that allows us to gain insights into subtle biases in existing data on foster youth, and to provide insight into where further (often qualitative) research is needed to identify potential ways of assisting youth. 

\end{abstract}
\section{Introduction}
There are over 420,000 children currently in foster care across the United States~\cite{afcars:2020}. 
Current and former Foster youth face a number of adverse outcomes in adolescence and early adulthood. For example, we know that of the roughly 25,000 foster youth who are never adopted or reunited with their families, 46\% are unemployed, one in four are homeless--- a rate around 200 times higher than the general population--- and one in three have dropped out of high school~\cite{Courtney:2011,Dworsky:2013,Oshima:2013}.

Scholars in the field of Social Work have spent decades identifying the factors that lead to poor life outcomes for foster youth, from systemic inequalities~\cite{race4,race3,gender} to funding challenges \cite{NBERw29906}. As in many social policy settings, one common source of data in these analyses are administrative data.  Specifically, myriad studies leverage the annually reported {\em Adoption and Foster Care Analysis and Reporting System} (AFCARS)~\cite{afcars:2019} foster care file, which contains individual-level data on foster youth across all 50 states, DC, and Puerto Rico who received services from government-funded agencies during that year.  However, there are a number of well-documented challenges that come with the use of such data \cite{CONNELLY20161}. In particular, they are 1) often missing critical information, 2) potentially difficult-to-work-with high dimensional categorical data, and 3) are biased by both systemic and individual-level factors \cite{race4,race3,gender,NBERw29906}.

In the present work, written jointly by computer scientists and social work scholars, the high-level \emph{technical} question is, \emph{(how) can we help Social Work scholars to use AFCARS data to help advance research that improves the lives of foster youth, while still accepting the shortcomings and difficultiees of the data?}
Our solution is to develop a novel clustering and cluster summarization approach that can be applied to high-dimensional categorical data to rapidly identify distinct and explainable clusters, or youth profiles, from coarse but large-scale administrative data. Our goal, then, is to use administrative data to \emph{inform future qualitative and/or experimental work}, rather than to try, as in most other technical work surrounding the foster care system, to make claims or predictions about youth based solely on lacking administrative records \cite{race4,race3,gender,Courtney:2011}.

More specifically, we propose an information-theoretic approach, using a mutual-information based scoring criteria to 1) identify and 2) summarize clusters. 
Our approach, unlike most other clustering methods for categorical data, does not require the number of clusters as an input, and also provides a novel approach to identify more easily explainable clusters. 
We evaluate our method in two ways. First, we show that the proposed method produces clustering performance superior to existing methods for categorical data~\cite{He:2008,avail} on a suite of benchmark data sets. Second, we conduct a case study in the utility of our method on foster care data from AFCARS in 2018. This case study, while brief, presents an example of how our method can be used to draw insights into real-world administrative data.

Our work, available \href{https://github.com/c4sgub/MIS_Categorical_Clustering.git}{here}, thus presents three primary contributions: 
\begin{itemize}[nolistsep]

    \item We propose a novel approach to clustering and cluster summarization for large-scale administrative data that outperforms state-of-the-art methods on benchmark datasets. 
    
    \item We identify novel and informative clusters of foster youth that we argue can help to shape future qualitative studies of foster care worker decision-making.
    
    \item Finally, we identify several systematic biases in the AFCARS dataset---the most widely used for studying foster youth---that warrant a careful consideration of which data are used, and how, if valid conclusions are to be drawn.
\end{itemize}
\section{Related Work}\label{sec:relwork}
%
The vast majority of data mining applications within the context of child welfare has focused on the use of predictive risk modeling. 
These models were designed, for example, to predict maltreatment substantiation~\cite{Vaithianathan:2013,Daley:2016,Rodriguez:2019}, 
 or to inform child welfare workers’ actions in response to screened-in maltreatment reports (e.g., removal, home-based support services)~\cite{Schwartz:2017}. Nearly all of these studies rely on administrative data of some kind, 
including the use of the datasets discussed here~\cite{Camasso:2019}. Our work offers a prescriptive, unsupervised method to help Social Work scholars understand potential patterns and data biases, rather than making (often biased) predictions about youth. 


To do so, we build on work focused on the clustering of categorical data. We do so because our data, and many other administrative datasets, are largely categorical in nature, and categorical data present unique challenges that have been addressed by these methods. Methods for clustering categorical data can be grouped into three  categories. Methods in the first category mimic the $k$-means algorithm by first randomly assigning the data instances into clusters and then iteratively redefining the clusters and reassigning the instances to the most appropriate cluster. COOLCAT~\cite{Barbara:2002}, $k$-ANMI~\cite{He:2008}, and G-ANMI~\cite{Deng:2010} are examples of this approach, and use information-theoretic measures to assign an instance to a cluster. However, they rely on knowledge of the optimal number of clusters. 
The second category of methods operates in a bottom-up agglomerative fashion, starting with individual data instances as clusters, and use a dissimilarity measure to recursively merge smaller clusters~\cite{Ganti:1999,Guha:2000}. For instance, CACTUS~\cite{Ganti:1999} uses the overlap between two attribute vectors, while ROCK~\cite{Guha:2000} uses the Jaccard coefficient. 

The method proposed in this paper falls in a third category of top-down agglomerative methods, which recursively split the data into partitions starting from a single cluster. The splitting process of this top-down approach can be used as an explanatory insight into the clustering process, which is a desirable feature for the domain analysts. 
Most similar to the present work is the MGR method~\cite{Qin:2014}, which selects an attribute with the maximum mean gain ratio and then chooses the partitions with the minimum entropy. Our work differs from MGR in the choice of the information theoretic measure. 

\section{Data}

Our analysis uses two types of data. First, in order to show that our method identifies meaningful clusters, we use seven publicly available and widely used data sets from the UCI repository \cite{Dua:2019}. 
We select a diverse array of data sets with varying sizes - from 101 to 12960 data samples, which have also been used as benchmark data sets by other methods to evaluate performance.  No changes are made to the data sets; even the samples with missing entries are used as-is.

Second, to show that our method has real-world utility, we conduct a case study on data from AFCARS. Although AFCARS is a national data set and all agencies are required to report on the same variables for all of the youth they serve, there are differences between states in how these variables are operationalized and recorded~\cite{Green:2015}. Here, we therefore focus our case study on data from two states that represent different models of child welfare administration: New York (NY) and Texas (TX)~\cite{Green:2015}. These two states represent a more decentralized and a more centralized approach to administration, respectively, and we thus expect them to differ in interesting and important ways.

The AFCARS data set contains over one hundred variables providing details about foster youth. In the present work, we restricted our analysis to a specific set of variables of theoretical interest to the Social Work scholars on our team. Specifically, our analysis included three sociodemographic, five clinical diagnostic, and 19 child welfare and family-related variables. Sociodemographic characteristics included sex
, race and ethnicity
, and a nine-category the rural-urban (R-U) continuum code representing the urbanization of the county in which the youth is located. 
We also included five dichotomous variables that captured whether or not the youth had been diagnosed as intellectually disabled, visually/hearing impaired, physically disabled, emotionally disturbed, or as having any other medical condition requiring special care. 

Child welfare-related variables included the manner in which youth were removed from their homes (voluntary, court-ordered, or not yet determined), whether parental rights had been terminated (yes or no), and the youth’s current placement setting (pre-adoptive home, relative foster family, non-relative foster family, group home, institution, supervised independent living, runaway, or trial home visit). The reasons for removal are separated into 15 dichotomous variables, each of which are coded as either applicable or non-applicable to the youth’s situation, full details on these variables are provided in our replication materials. Finally, we included one variable describing the structure of the family from which the youth was removed.

Administrative data often suffers from missing data problem, AFCARS data is no exception. Commonly used methods to handle missing data are data imputation techniques such as mean substitution, regression imputation, maximum likelihood~\cite{doi:10.1080/08839514.2019.1637138}. These methods require making parametric assumptions regarding data generating process; which for our purpose of analysis isn't required as the task in hand is to study the data itself rather than using data for downstream tasks like prediction. We impute the missing data with a separate missing data category labeled '?'. This has key advantages; 1) does not require any parametric assumptions 2) provides us a way to uncover, if any, non-random missing data. 

\section{Method}

\subsection{MIS Clustering Method}
\label{mis:method}
Our approach is a top-down clustering method that clusters the data using a mutual information-based scoring metric. Formally, our goal is, given a set of data samples $O=\{o_1,...,o_n\}$ (e.g. foster youth), described by a set of attributes $A=\{a_1,...,a_r\}$ (sociodemographics, etc.), to partition $O$ into a set of clusters $C=\{C_1,...,C_k\}$ such that the samples (youth) within each cluster 1) share at least one attribute and 2) are similar to one another.  We argue (and show) that this leads to effective, interpretable clusters. Note that each attribute is characterized by two or more \emph{categories} (e.g. the attribute placement setting has categories pre-adoptive home, group home, etc.). 



Our algorithm recursively creates clusters via a two step procedure. First, it identifies a \emph{significant attribute}, which we define intuitively as the attribute that \emph{provides the most information about the structure of the data to be clustered}. To identify the significant attribute, we must define a measure of which attribute provides the ``most information.'' We do so using a modified mutual information score. We first define mutual information:

\begin{definition}[Mutual Information]
For attributes $a_i,a_j \in A$ with domain sizes (number of categories in an attribute) of $l$ and $m$ respectively, and which define a partition $O/a_i=\{P_1,\cdots,P_l\}$ and $O/a_j=\{Q_1,\cdots,Q_m\}$ respectively on $O$, the mutual information between these two attributes is written as follows, where the probability $P(P_s)= \frac{\vert P_s\vert}{\vert O\vert}$ and the joint probability $P(P_s, O_t)=\frac{|P_s \cup Q_t|}{\vert O\vert}$; $P_s, Q_t \subseteq O$.
\begin{equation}
    MI(a_j,a_i) = \sum_{s=1}^{l}\sum_{t=1}^{m} P(P_s, Q_t)\log_2\frac{P(P_s, Q_t)}{P(P_s)P(Q_t)},
\end{equation}
\end{definition}

Using this definition, we then define the \emph{mutual information score} (MIS) of each attribute as follows, where $l \vert O/a_i \vert$ is the number of partitions defined by $a_i$ on $O$ also referred to as domain size of $a_i$:

\begin{definition}[Mutual Information Score]
For an attribute $a_i \in A$  which defines a set of partition $O/a_i=\{P_1,\cdots,P_l\}$ on $O$. The mutual information score is defined as 
\begin{equation}
    MIS(a_i) = \frac{\sum_{j=1, j\neq i}^{\vert A\vert} MI(a_j,a_i)}{\vert l \vert},
    \label{mis}
\end{equation}
\end{definition}

Note that in the definition of MIS above, the standard definition of mutual information is divided by the number of partitions defined by significant attribute. We do so in order to offset known biases in mutual information, where mutual information is generally greater for attributes with more categories and lower for fewer data samples~\cite{pmlr-v32-romano14}. Bias towards fewer data samples does not affect our method as we compare attribute columns, and each of these columns has the same number of samples. However, we do need to offset the bias introduced due to the differences in the number of categories in each attribute.


Having identified the significant attribute, we then create data partitions based on categories of the significant attribute. For example, if the significant attribute was \emph{manner in which youth was removed}, partitions are created based on its categories: voluntary, court-ordered, or not yet determined.
Data samples that are similar when grouped together result in low entropy \cite{Barbara:2002}. Thus, the partition with the least entropy is selected to form a new cluster, $P_i$. The entropy $H$ of a partition $P_i$ can be written as the joint entropy of set of attributes $A=\{a_1, \cdots ,a_r\}$, that is, as $H(P_i) = H(a_1,\cdots,a_r) = \sum_{a \in A}H(a)$ if and only if attributes are statistically independent. Independence of the attributes cannot always be guaranteed and therefore our measure of \emph{partition entropy} is rather an approximation, defined as:

\begin{definition}[Partition Entropy]
Given set of attributes $A = \{a_i, \cdots ,a_r\}$ and a partition $O/a_i = \{P_1, \cdots ,P_l\}$ induced by a significant attribute $a_i \in A$ 
\begin{equation}
  H(P_i) = \sum_{i=0}^{r} \sum_{x \in P_i} -\ p(x)\log_2p(x)  
\end{equation}
\label{partition_ent}
\end{definition}

\subsection{Cluster Summarization}
Our MIS clustering approach identifies clusters that share a single attribute and are similar along other attributes. Initial use of the tool with Social Work scholars suggested, however, that it would be most useful if we also were able to explain, or summarize, \emph{how these clusters were similar}. To do so, we construct a method based on KL-divergence. Specifically, let $A=\{a_1,\cdots,a_k\}$ be the set of attributes associated with all the data samples $O$, we refer to this as \emph{global attributes}. Let $A^i_c=\{a^i_1,\cdots,a^i_k\}$ be the set of attributes associated with data samples belonging to the cluster $p_i \in P$, where $P = \{p_i,\cdots,p_k\}$.
We measure the KL-divergence between the probability distribution $q_j$,$p_j$ of the cluster attributes $a^i_j$ and $a_j$; with a set of states $X$ and global attributes as
\begin{equation}
    D(q_j,p_j) = \sum_{x \in X}q_j(x)\log_2\frac{q_j(x)}{p_j(x)},
\end{equation}


\section{Comparison with Other Methods}\label{sec:evaluation}

Table \ref{accuracy} shows that our MIS algorithm either outperforms or has comparable performance to other state-of-the-art methods on 4 out of 6 standard data sets from the UCI repository.  We compare our proposed method to five other state of the art categorical clustering methods methods introduced in Section~\ref{sec:relwork}: MMR, MGR, k-ANMI, G-ANMI, COOLCAT, and $K$-modes. The $K$-modes algorithm was evaluated using an available implementation~\cite{devos2015}, and the results for the remaining methods are reported from the original papers. Finally, our proposed algorithm MIS can operate with or without providing the number of clusters. In  order to make fair comparisons, we set the number of clusters to the number of real classes for the respective data set, similar to the evaluation of the other methods we analyze. We also have provided results for clusters obtained without specifying number of cluster as MIS-auto.  

We use purity to evaluate the performance of each method. Purity is an external evaluation metric that measures the extent to which a cluster overlaps with a class. For a set of clusters $C=\{C_1,\cdots,C_k\}$ and classes $D = \{D_1,\cdots,D_d\}$, purity is defined as $\frac{\sum_{i = 1}^{k} \max_{j = 1}^{d} \mid C_i \cap D_j\mid}{N}$, where $N$ is the total number of data samples. Purity is bounded between 0 and 1, wherein 1 indicates perfect clustering, i.e all data samples in a cluster belong to the same class.

MIS performs exceedingly well on the Mushroom data set, which contains 22 attributes describing each of the 8124 mushrooms. Out of 22 attributes, `odor', which has 9 different categories, is the attribute with the highest total mutual information (MI). However, since the MI is artificially boosted for attributes with greater domain size, MIS counters this and determines `bruises' to be a more suitable significant attribute.
MIS's performance is on par with MMR and MGR on the Balance data set and otherwise outperforms these methods. G-ANMI has the best purity score for the Vote data set when the number of clusters is specified, however, MIS-auto outperforms G-ANMI. 
In general, the performance of MIS-auto is greater than or equal to performance of MIS with specified number of clusters, which in part is due to the bias discussed in Section~\ref{mis:method}.

\begin{table}[t]
\small
\centering
\caption{Purity of categorical clustering algorithms on UCI data sets.}
\label{accuracy}
\resizebox{\linewidth}{!}{
\begin{tabular}{l l l l l l l | l}
\toprule
    \textbf{Algorithm} & \textbf{Zoo} & \textbf{Vote} & \textbf{Cancer} & \textbf{Mushroom} & \textbf{Balance} & \textbf{Chess} & \textbf{Average}\\  
\midrule
\textbf{MGR} & \textbf{0.930} & 0.827 & 0.864 & 0.677 & \textbf{0.635} & 0.533 & 0.744 \\
        \textbf{MMR} & 0.911 & 0.687 & 0.669 & 0.518 & \textbf{0.635} & 0.523  & 0.657\\
w        \textbf{K-MODES} & 0.860 & 0.852 & 0.651 & 0.560 & 0.587 & 0.503  & 0.668\\
        \textbf{k-ANMI} & 0.733 & 0.869 & \textbf{0.978} & 0.587 & 0.506 & 0.547  & 0.703 \\
        \textbf{G-ANMI} & 0.874 & 0.871 & 0.966 & 0.547 & 0.518 & 0.543 & 0.719\\
        \textbf{COOLCAT} & 0.785 & 0.839 & 0.650 & 0.531 & 0.506 & 0.533  & 0.640\\
        \midrule
        \textbf{MIS} & 0.891 & 0.828 & 0.882 & 0.743 & \textbf{0.635} & 0.533  & 0.752 \\
        \textbf{MIS (auto)} & 0.891 & \textbf{0.949} & 0.927 & \textbf{0.828} & \textbf{0.635} & \textbf{0.558} & \textbf{0.80}\\
        
    \bottomrule
\end{tabular}
}
\end{table}

\section{Case Study}

We applied our MIS algorithm and cluster summarization approach to AFCARS data for youth in New York (N=23,676), resulting in 10 clusters, and in Texas (N=52363), resulting in 6 clusters. As is typical in unsupervised modeling, some clusters offered clear insights, others did not. This brief case study is organized around three main insights that were gleaned via analyses of cluster summaries produced by our method by Social Work scholars:


\textbf{1. Clear patterns of non-randomness in (non-)missing data:} Many of the clusters in our data
were, surprisingly, largely defined by the \emph{absence of missing values}. That is, the salient factor which differentiated these clusters from all others were that they had significantly more complete data on certain attributes than one would expect by chance.  The high percentage of missing values overall is not unexpected in administrative data. However, the patterns our clustering algorithm identifies in where data was \emph{not} missing offered our team new insights into the nature of \emph{how} data were missing, and thus informed our understanding of the ways in which data seem to have been collected. 

For example, 
we identified two clusters of youth in New York which had both  a) no youth with missing values for various Clinical Diagnosis attributes (e.g. ``Clinically Diagnosed with an Emotional Disability''), compared to a base rate of around 12\% in the general population, and b) were heavily characterized by particular Placement Settings.  In one of the clusters, 69\% had a Placement Setting of {\it Pre-adoption home}, meaning a home into which they were likely to be adopted, compared to only 12\% of all youth. And in the other, youth were almost twice as likely as the base rate to be in a Foster Care setting. These findings suggest differences in the accessibility or completeness of information about youths’ medical histories across different placement settings. 

These non-random patterns of missing values are particularly critical because they vary along youth Placement Setting, perhaps the most important variable in understanding the trajectory of a youth through the foster care system \cite{connell2006changes}. While such patterns of missing values can potentially be remedied, our analysis presents the first evidence that we are aware of to identify these non-random patterns of missing values in the widely-used AFCARS data set.

{\bf 2. The importance of viewing the data that represent youth holistically:} Because our cluster summarization approach allows us to construct profiles of youth that are unique (from a mutual information perspective) across many attributes, we are able to better study more general patterns of differences across profiles of youth rather than focusing on differences in specific levels of specific attributes. For example, in Texas, we identified one cluster representing a small subset of children in voluntary placements. High percentage of these youth were in trial homes (22\%) and relative foster care (36\%). The number of youths placed back into their homes in this cluster skewed lower relative to the overall sub-sample. In contrast, a second cluster in Texas had significantly higher percentages of youth in pre-adoptive placements (40\%) and non-relative foster homes (30\%) and placement disruptions that skewed higher relative to the overall sub-sample. 

The implication of this is that there is an inextricable relationship between these different kinds of placement types and the extent to which a youth ``bounces around'' in the foster care system. There are many possible reasons why this linkage between placement types and number of placement settings might exist; for example, youth who have been in care for longer periods of time often experience many placement disruptions and lose connections to relatives. However, to the best of our knowledge, this linkage has not been previously identified in the literature, thus showing the utility of our method in identify new pathways for future work. 


{\bf 3. State-level funding decisions may have influenced the structure of the clustering results, at least in New York} Some clusters we identified seemed to reflect patterns related to different funding eligibility criteria, which overlaps with the placement type attribute. For example, we identified one cluster of youth in New York who were predominantly in pre-adoptive placements (23\%) or relative foster homes (44\%). There were far fewer youth in non-relative foster homes (3\%), group homes (12\%), supervised independent living programs (14\%), and institutions (4\%). Some of these placements may not have been approved as licensed foster homes \cite{nys_child_2018}. Youth may live with a relative who does not have legal custody for several months before the relative petitions the court and becomes a certified foster caregiver or pursues adoption, or may have a criminal history or safety issue in the home that precludes licensure.  

These youth, and those with whom they are placed,  might therefore not have been eligible for certain services or subsidies, which in turn may have influenced their outcomes that are reflected in AFCARS. These clusters that seem to be driven in part by the ways in which state policies revolve around funding decisions suggest critical future work in understanding the relationship between state-level policy and administrative data.
\section{Conclusions}
We have described a novel clustering algorithm for categorical data which uses an information-theoretic splitting criterion. The algorithm is significantly better (See Table~\ref{accuracy}) than other state of art algorithms on several benchmark data sets. At the same time, the KL-divergence based interpretability strategy offers an explainable summary of the clusters, which is a highly desirable feature when presenting the results to domain researchers. In particular, the algorithm, when applied to the AFCARS data, revealed new potential insights that suggest the need for further (social) theory, and both qualitative and quantitative work into better understanding the impact of the youth's characteristic on outcomes.

However, it is crucial to remember, as we begin to apply machine learning to high-stakes child welfare decision-making, that tools like this clustering exercise can aid in understanding, and perhaps help guide policy and practice decisions, but data always tells an incomplete story. Even if a child is well-represented by clustered attributes, personal knowledge of the child will always be important when making decisions about that child’s needs. 
%
%
%
\bibliographystyle{splncs04}
%

\bibliography{refs}





\end{document}